# An ASIC Implementation and Evaluation of a Profiled Low-Energy Instruction Set Architecture Extension

Bobby Sleeba, Mikael Collin and Mats Brorsson

*KTH Microelectronics and Information Technology,*
*Royal Institute of Technology, Stockholm, Sweden*

{sm02-slb, mikaelc, Mats.Brorsson }@imit.kth.se

**Abstract**

This paper presents an extension to an existing instruction set architecture, which gains considerable reduction in power consumption. The reduction in power consumption is achieved through coding of the most commonly executed instructions in a short format done by the compiler based on a profile of previous executions. This leads to fewer accesses to the instruction cache and that more instructions can fit in the cache. As a secondary effect, this turned out to be very beneficial in terms of power. Another major advantage, which is the main concern of this paper is the reduction in the number of instruction fetch cycles which will also contribute significantly towards reduction in power consumption. The work involves implementing the new processor architecture in ASIC and estimation of power-consumption compared to the normal architecture

## 1 Introduction

Embedded computer systems account for most of the computers in the world. The performance requirements in these systems are rising rapidly and consequently, the power consumption have also become an important issue since high performance systems also consume a great deal of power. Previous studies have shown that the instruction fetch path is a major source of energy consumption in computer systems targeted for embedded systems. In [4] we showed that about 25% of the energy consumed stems from the instruction fetch path. Of this, about 25% is in the branch prediction logic and 75% in the instruction cache. A similar percentage of the execution time can be attributed to the instruction fetch path.

In [4], we also proposed an extension to a normal RISC instruction set architecture (ISA) to reduce the energy consumption due to instruction fetches in embedded applications. It is based on a profiled extension to the normal ISA where the most commonly executed instructions, as determined by the profile, are encoded as short instructions, thus occupying less space in memory but more importantly also in the cache. Decoding is done through lookup tables where the short instructions have bit fields, which are indexes to the lookup tables, which in turn contain the opcode and arguments of the full-length instruction.

In order to make the table lookup we have introduced a new pipeline stage between the standard fetch and decode stages that many pipelined processors have. Our design affects only the fetch stage and the new stage, which we call depack. The decode stage, and the following stages, are unaffected. Architecture-level simulations have shown that we can achieve up to 15% reduction in total energy consumption on a selection of benchmarks.

In [4], we outlined the design of our compressed instruction decoder. The contribution of this paper is the presentation of a full design and a VHDL implementation of our compressed instruction set decoder.

## 2 Related work

The use of short instructions to improve the usage of the instruction cache memory and the memory bandwidth is not new. Both ARM [12] and MIPS 16 [8] have 16-bit variants of their normal 32-bit instruction sizes. However, the short instruction sets are not as powerful as the normal instruction sets and require more dynamically executed instructions and extra transitions between 32-bit and 16-bit modes.

Stan and Burleson describe a coding scheme by Bus Invert [7]. If the Hamming distance of two consecutive data words is greater than half of the word size, the inverted data is send and a control signal is asserted. A theoretical power saving of 25% can be done on average. In their solution, there is no change in fetch bandwidth or code size. In addition, the extra control signal will be a compromise for the power saving. In [5], an encoding scheme is introduced, where the most frequently used consecutive instruction pairs have their opcodes with the smallest Hamming distance possible which reduces the switching activity from 37% - 67%. The paper introduces a new encoding scheme, which reduces power but the code size and fetch bandwidth are not affected here.

Other approaches have been to store the program compressed in memory and to decompress the instructions dynamically into the instructions cache [9][13]. This reduces the storage for instruction memory space and the number of bytes fetched from memory, but it does not reduce the instruction cache memory bandwidth. A similar approach is used in Pentium 4, which stores traces of decoded microinstructions in an internal trace cache that replaces the normal level-one instruction cache [6].

In [11], a run-time reconfigurable mechanism to map multiple instructions in a single compressed bit pattern is proposed. This reduces the flexibility of compression obtained by variable length instruction encoding.

Another interesting approach having variable length instructions was done with a similar aim in [10] although they did not consider any execution-based evaluation. The approach in [1] also uses a lookup table for decompression. However, the decompression unit is inserted between the processor and the memory, which increases the critical path. Their scheme is not integrated with the processor pipeline, which makes it easier to use with proprietary processor models. However, this also might lead to problems when integrating it with the memory hierarchy and in particular with branch address targets and calculations, which are done in a different address space.

## 3 A profiled variable-length instruction set

The general idea that we are proposing is to encode the most commonly executed instruction in a short format and then augment the processor with the ability to expand them to normal 32-bit instructions inside the front-end of the pipeline. Thereby we should gain performance by being able to fit more instructions in the cache and power by reducing the number of accesses to the instruction cache as often as before.

We started with the standard MIPS instruction set and changed the encoding of instructions so as to create space for short instructions starting with a 0 in the most significant position. We profiled a number of SPEC CPU2000 benchmarks and found that there was not only locality in which instructions that were executed, but also in which arguments they used. Some arguments were more common than others. Although the SPEC programs are not the most common programs to be run in embedded systems, they provide us with the insight needed for this evaluation.

### 3.1 Short instructions

Our proposal is to use a short, variable length, instruction set in parallel with the normal 32-bit instructions as depicted in Figure 1. Based on the profile measured, we have selected five instructions to be encoded as 8-bit instructions and two to be encoded as 16-bit instructions using table lookup for the arguments. The opcode is encoded with 2-4 bits and this selects a set of lookup tables for the arguments. The ADDIU-instruction, for example, activates two argument lookup tables, one with 16 elements to select the register combination arguments of this instruction, and one table with eight elements to select the immediate argument. The different sizes of the tables are determined by the locality in arguments. For the LW-instruction, the use of the immediate argument has less locality than the register arguments. Therefore the argument lookup table for immediate values is bigger. For the ADDIU-instruction the opposite is true. The total coverage of this coding scheme can be found by adding the relative frequencies of usage, which can be seen in the table above. In total it becomes 50.6% plus the coverage for a few other instructions, which can be coded with 16-24 bits without the need for lookup tables.

| Instruction | Instruction Coding | | | | | | | Frequency of Usage |
|---|---|---|---|---|---|---|---|---|
| LW    | 0 | 1 | 0 |   | R |   | IMM | 10.40% |
| ADDIU | 0 | 1 | 1 |   | R |   | IMM | 4.53% |
| SW    | 0 | 0 | 1 | 1 | R |   | IMM | 2.25% |
| SLL   | 0 | 0 | 1 | 0 | R |   | IMM | 5.41% |
| ADDU  | 0 | 0 | 1 | 1 |   | R |     | 21.93% |
| BEQ   | 0 | 0 | 0 | 0 | 0 | R | 8-bit IMM | 4.06% |
| BNE   | 0 | 0 | 0 | 0 | 1 | R | 8-bit IMM | 2.04% |

**Figure 1. The coding of the most frequent instructions.**

### 3.2 Architecture implications

Figure 2 shows the general structure of the front-end of a processor that we are considering supporting the encoding scheme previously described. The fetch stage fetches *chunks* of four bytes from the instruction cache. A chunk may contain zero to four instructions depending on the type of encoded instruction. The fetched instruction(s) are then put into a ring buffer called *depack_Q*. This queue effectively acts as a pipeline register separating the fetch stage with the added *depack* stage (DP) which extracts one instruction of length 8, 16, 24 or 32 bits from the depack_Q and, using the argument lookup tables, to recreate and insert a full-length 32-bit instruction into the register separating the depack and decode stages. From here on the pipeline remains unaltered performing decode and execution of normal uniform length RISC style instructions.

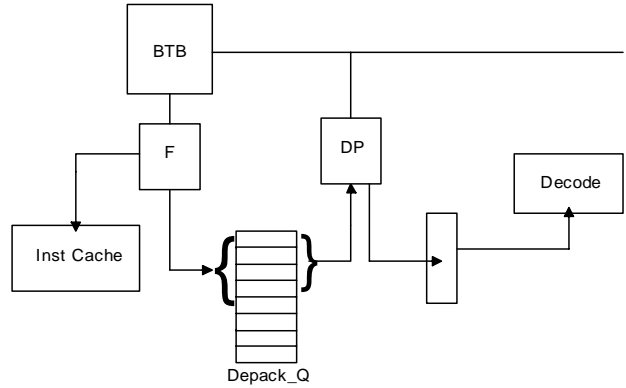

**Figure 2. The general structure of the micro-architecture needed to accommodate the variable instruction set.**

The branch prediction logic is normally checked on every instruction fetch. As we are not fetching instructions, but chunks, which may contain more than one instruction, we made the decision to check the branch prediction logic on chunk fetches. This forced us to make the restriction that we allow only one branch to start in every chunk. This will be the branch, if any, that we will predict. In the case when a branch is predicted taken, but we have not fetched the full instruction yet as it might cross a chunk boundary, the fetch and depack stages must work together to continue sequential fetch of the next chunk although we predict the fetching of a new chunk.

Architecture simulations using four SPEC CPU2000 programs show that we can gain up to 15% in total energy consumption using the scheme depicted above as shown in Figure 3 below [4]. Although these simulations are based on a detailed model of the behavior and power consumption, we felt that we needed a more hands-on experience in the difficulties in implementing our approach, which is why we looked into implementing the critical stages of the pipeline front-end in an ASIC deisgn. This implementation is discussed and presented in the next section.

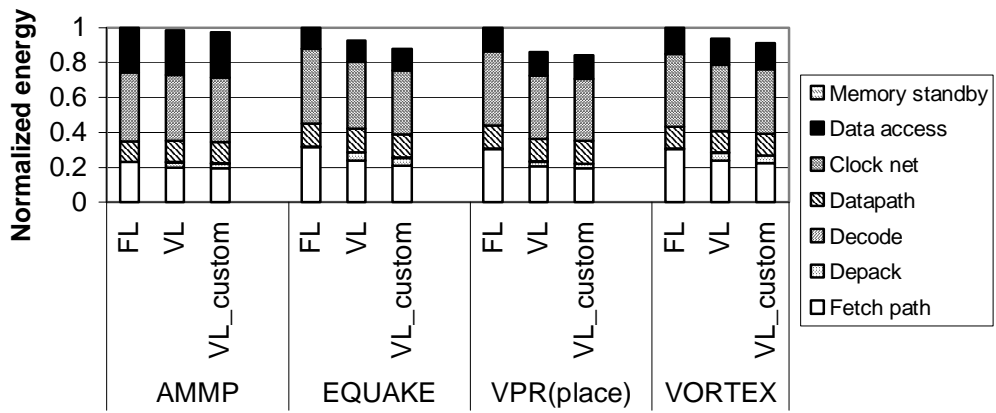

**Figure 3. Breakdown in energy for four different applications relative the fixed-length instruction set.**

# 4 Proposed micro-architecture implementation

We have implemented the pipeline front-end in VHDL and synthesized the design using the UMC 0.18μm standard-cell library. The overall structure of the new front-end is shown in Figure 4, which is very similar the general structure of Figure 2. In addition to the fetch stage, for our purposes moderately modified to fetch instructions of variable length, the major issue is the insertion of the Depack stage.

The fetch stage is responsible for fetching the next chunk of instructions, in sequence or from the address given by the branch prediction logic. The word-aligned chunk address is presented to the instruction cache, which responds by inserting the fetched chunk in to one of the 32-bit wide pipeline registers denoted: **A** & **B** in Figure 4, or stalls on cache miss. The registers **A** & **B** constitutes the previously mentioned Depack_Q. The task of the new depack stage is to extract one to four bytes from the Depack_Q in sequence, to decompress and then forward them to the normal decode stage. The length of short and normal instructions is calculated and arguments are retrieved using lookup in argument tables. The branch prediction logic outputs the predicted branch target address and asserts the *Branch control* signal to both Fetch and Depack stages respectively, which serves as a control signal in case a branch is predicted taken or if execution is redirected due to a previous miss prediction.

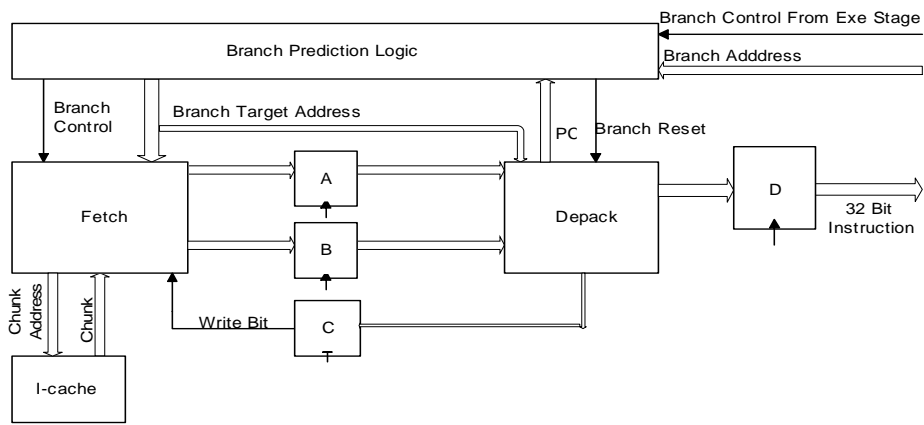

**Figure 4. The top-level design of the fetch and depack stages and the branch prediction logic.**

## 4.1 The fetch stage

A detailed description of the modified fetch stage is shown in Figure 5. The next 32-bit chunk to fetch is pointed out by the *CC* register[1]. The CC register contains the updated value of the chunk address, i.e., the next in sequence or the branch target address, controlled by the Branch Control signal. Data retrieved from the cache is inserted in to one of the two registers **A** or **B** selected through clock gating using the signals *Write Enable 1*, and *Write Enable 2*. Information to the *Fetch Control* FSM, for selection of which register to update with the fetched chunk is provided by the Depack stage through the *Write_ bit*, and Branch Control signals. Initially, on reset, and when branches are taken, register **A** is selected. The actual fetch, presenting the word-aligned chunk address is controlled by the *Fetch*

---

[1] CC Chunk Counter, analog to PC Program Counter.

*Enable* signal, which stalls the fetch stage when there is no space for the chunk to be fetched in any of the Depack_Q registers. Fetch also stalls on a cache miss.

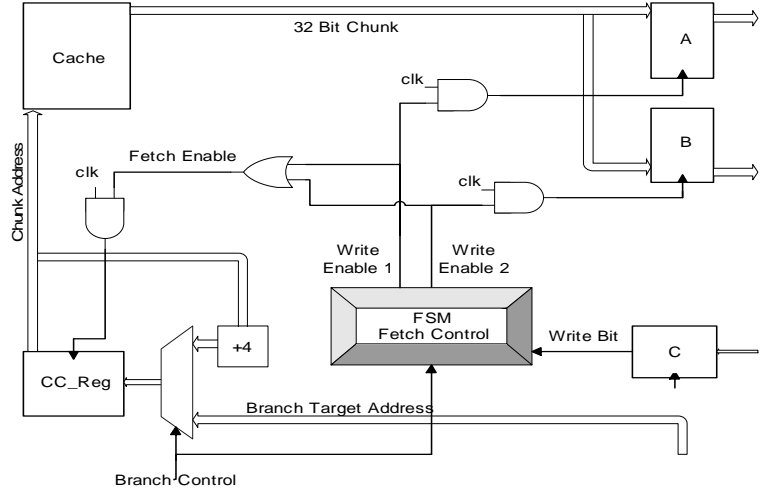

**Figure 5. Implementation of the fetch stage.**

## 4.2 The Depack stage

The De-Pack stage shown in Figure 6 is responsible for decompressing fetched chunks, containing encoded variable length instructions, that are in the Depack_Q into 32 bit instructions. The Depack_Q is implemented as a ring buffer in which each of its eight bytes is individually addressable through a designated *Read Pointer* (RP). Data stored in the Depack_Q registers are fed to the *Depack Logic* using four multiplexers. Each one selects one of the stored 8 bytes using the Read Pointer. Four consecutive addresses are supplied to the multiplexers such that the Depack Logic gets four adjacent bytes of data starting from the byte address pointed to by the Read Pointer generated by the *Read Control Logic* FSM.

The Depack Logic uses a number of bits in the first of the four extracted bytes to decode instruction type and length in number of bytes. The length information is then fed to the Read Control FSM, which calculates the Read Pointer for next cycle. When a branch occurs, the Read Pointer is updated with the two least significant bits of the byte aligned branch target address. Hence the Read Pointer fills the gap between chunk addressing and byte addressing by the branch address. The most significant bit of the Read Pointer is also fed back to the Fetch stage as the Write bit, controlling which register to insert the fetched chunk in.

The Depack Logic gets decoding information for register combinations and eventual immediate values from the *Look up Tables* (LUTs). The instruction is then transformed into the format of the normal uniform length instructions, with its original opcode regenerated, and register combination and possible immediate value inserted from the LUTs. The complete instruction is then written to the pipeline register separating Depack and Decode. The LUT was implemented as a ROM.

The branch prediction logic is given a byte-aligned Program Counter (PC), pointing to the next instruction to Depack. This new PC value is calculated using length information retrieved from the actual depacked instruction.

When encoding the short instruction's opcodes, care was taken to reduce the number of identification bits to use for the encoded instructions when compared to the normal 32-bit instructions. This technique will result in faster decode of the compressed instructions, but will not impose any disadvantage for the 32 bit instructions, as no processing is needed to Depack them. Hence, we were able to optimize the Depack time for the variable length instructions.

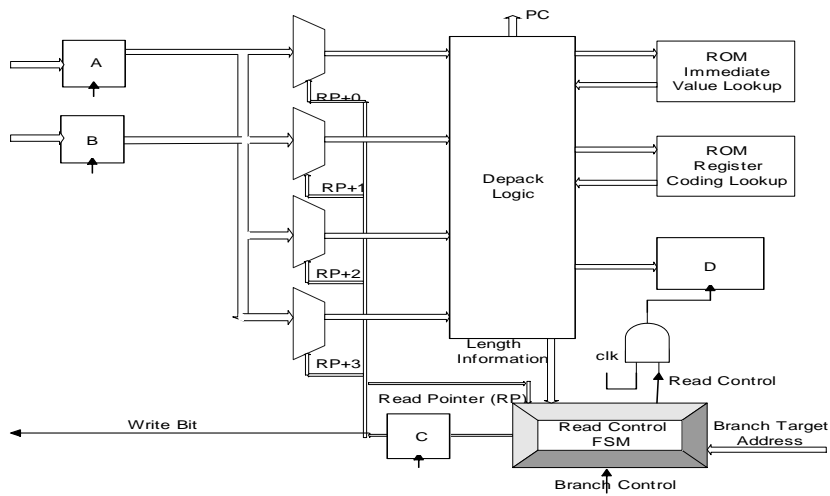

**Figure 6. The Depack stage generating the read address and 32-bit instructions.**

## 5   Simulation, Synthesis and Power Estimation

Hardware was designed in VHDL, simulated in Modelsim-5.7, synthesized in Synopsys using the UMC18 - 0.18μm 1.8V library. Power estimation was done using both Synopsys and Modelsim-5.7, and area estimation was done in Synopsys Design Analyzer.

Powers estimation was done using both RTL and gate level power estimation methods. For getting RTL Path dependent power the first step was to create a SAIF forward annotation file (having all the technology independent signals) using design analyzer. In the case of gate level power estimation, technology dependent signals were taken. This file was linked to Modelsim-5.7 by Synopsys foreign language interface. In the second step, a test bench, which inputs instructions according to the percentage of their occurrence, was run in Modelsim-5.7. The information of switching activity during the simulation was outputted to SAIF back annotation file, which served as input to Synopsys during its power estimation. Power was estimated considering the circuit operates at 300 MHz with operating condition as worst case and with wire load model 20k.

As the simulation of the entire standard applications was not practical for hardware simulation, instructions were organized according to the percentage of occurrence, obtained from the SimpleScalar simulator [3] and were the fed as input to the simulator.

## 6   Results

The aim for this work has been to explore the design to get a feel for the complexity involved when enabling the pipeline front-end with the ability to fetch and depack instructions of variable length. The complexity we wanted to explore was the extra power imposed by the modified fetch and depack stages. We were also interested in the effects this had on area and timing.

## 6.1 Power Comparison

Total dynamic power is contributed by both cell internal power and net switching power. Gate level power comparison is done in Table 1. To the left of the double line we show numbers for our modified pipeline front-end and to the right we show the numbers for a normal fetch stage. The instruction cache or the branch target buffer are not included.

**Table 1. Gate level power comparison.**

| GATE level power | FETCH | | DEPACK | | TOTAL (front end) | | Normal architecture | |
|---|---|---|---|---|---|---|---|---|
| | mW | % | mW | % | mW | % | mW | % |
| Cell internal power | 7.957 | 78.8 | 5.048 | 63.0 | 13.01 | 72.2 | 1.74 | 72.3 |
| Net switching power | 2.134 | 21.2 | 2.929 | 38.0 | 5.063 | 27.8 | 0.662 | 27.7 |
| Total dynamic power | 10.09 | 100 | 7.977 | 100 | 18.07 | 100 | 2.404 | 100 |
| Cell leakage power | 0.164 mW | | 0.048 mW | | 0.212 mW | | 0.029 mW | |

From Table 2, which compares the RTL and gate level power for different stages in the design we can find that the Fetch stage consumes considerably more power than the Depack stage. This is mainly due to the depack_Q which is referred to the fetch stage and which accounts for much of the switching power. Considering the powers for normal full length and variable length architecture we could note a 7.5 times increase in power consumption for the additional hardware introduced compared to the normal architecture.

**Table 2. A comparison of the power consumption in the fetch and depack stages in the two architectures.**

| New architecture | Total RTL power | | Total gate level power | |
|---|---|---|---|---|
| | mW | % | mW | % |
| Fetch | 9.86 | 61 | 10.09 | 74.4 |
| Depack | 6.94 | 39 | 4.64 | 25.6 |
| Total | 17.79 | 100 | 18.23 | 100 |
| Normal architecture | 1.11 | | 2.43 | |

As the remaining stages and memory structures of the total processor was not implemented during this study, the power consumption for those parts was estimated using the Wattch [2] framework. Wattch was configured to calculate the power for those structures using parameters corresponding to a generalized 0.18μm process provided by Wattch. The power figures obtained from Wattch was then scaled to the power of the depack stage obtained from Synopsys. The scaling factor was used to scale the total power for the complete processor. When comparing the two architectures regarding this estimated power consumptions we found that the modified Fetch, and the added Depack stages, only contributes with an increase of 2.17% to the total processor power. As shown in [4], however, the added power consumption in the pipeline is compensated with less cache and branch prediction buffer accesses, leading to a reduction in energy consumption.

## 6.2 Critical Path

The critical path was found to be in decoding the Read Pointer from the registers A & B. This was because 8 bits of data read from the registers pointed by the Read Pointer will have to be checked to find the length of the instruction and this data has to be added with the previous Read Pointer value to obtain the new Read Pointer. It was also noted that there was only small difference between the critical paths due to a 32-bit addition of the PC value with 4 in the normal fetch stage. Hence we could conclude that the critical path is not affected in the new architecture. The difference in maximum frequency for the two architectures is thus negligible. The critical path could be reduced by reducing the number of encoded bits to check the length of instruction, which can be done by optimizing the instruction set architecture for minimum length decoding.

## 6.3 Area Comparison

The area was obtained as shown in Table 3 with Normal and modified architecture. We can note from the table that with the additional Depack stage and the modified fetch stage, the area for the new hardware has increased by 3 times the area for the normal fetch stage. Another interesting thing to note is that the combinational area is increased 8.7 times while the non-combinational area has increased only 1.6 times compared to the normal fetch architecture. The combinational area is noted to be increased considerably in the Depack stage mainly due to four large multiplexers used to multiplex 8 byte data from registers A and B.

**Table 3. A comparison of combinational and Non-combinational area for the Normal and New architecture.**

| AREA | New-Fetch | | Depack | | Total | | Normal | |
|---|---|---|---|---|---|---|---|---|
| | $(10^3)$ | % | $(10^3)$ | % | $(10^3)$ | % | $(10^3)$ | % |
| Combinational | 4.475 | 39 | 11.002 | 78.1 | 15.477 | 54 | 1.792 | 19 |
| Non Combinational | 8.278 | 61 | 3.089 | 21.9 | 12.168 | 44 | 7.415 | 81 |
| Total cell area | 13.56 | 100 | 14.091 | 100 | 27.646 | 100 | 9.208 | 100 |

## 7 Conclusions

Our goal to analyze the complexity of implementing the pipeline front-end to fetch and depack instructions of variable length has shown that no reduction in clock frequency is needed. The area of the new front-end is increased with a factor of 3, which when considering the whole processor with caches and logic, it is not that alarming. When looking at the two front-ends of the two architectures regarding gate level power we see that the new architecture is needs 7 times more power than the full instruction-length architecture. This is not as critical as it first appears when considering the total processor power consumption.

Our estimations of the impact on total processor power consumptions is an increase of 2.17%, indicating that it has a rather small impact. Since the usage of variable length instructions reduces the number of cache accesses the total amount of energy consumed during runtime is also reduced.

**References**

[1] L. Benini, A. Macii, E. Macii, and M. Poncino, "Selective Instruction compression for Memory Energy Reduction in Embedded Systems", in *Proceedings of 1999 International Symposium on Low Power Electronics and Design*, 1999, pp. 206-211.

[2] D. Brooks, V. Tiwari V., and M. Martonosi, Wattch: A Framework for Architectural-Level Power Analysis and Optimizations, in *Proceedings of the 27th International Symposium on Computer Architecture, ISCA'00*, June 2000, pp. 83-94.

[3] D. Burger and T. M. Austin, *The SimpleScalar Tool Set, Version 2.0*, Computer Architecture News, June 1997, pp. 13-25.

[4] M. Collin and M. Brorsson. "Low Power Instruction Fetch using Profiled Variable Length Instructions", in *Proceedings of the IEEE International SoC Conference*, Sept. 17-20, Portland, Oregon, 2003.

[5] S. Kim and J. Kim, "Opcode encoding for low-power instruction fetch", *Electronics Letters, Volume: 35 Issue: 13*, 24 June 1999, pp.1064 -1065.

[6] P. N. Glaskowsky, "Pentium (Partially) Previewed", *Microprocessor report*, {8/28/00-01}, August 28, 2000, pp.10-13.

[7] M. R. Stan, and W. P. Burleson. "Bus Invert Coding for low power I/O ", *Very Large Scale Integration (VLSI) Systems, IEEE Transactions on , Volume: 3 Issue: 1* , March 1995 , pp.49-58

[8] K.D. Kissell. "MIPS 16: High-Density MIPS for the Embedded Market", in *Proceedings of Real Time Systems ´97 (RTS97)*, 1997.

[9] H. Lekastsas, J. Henkel, and W. Wolf, "Code compression for Embedded System Design", in *Proceedings of the 37th Design Automation Conference*, June 2000, pp.516-521.

[10] H. Pan and K. Asanovic, "Heads and Tails: A Variable Length Instruction Format Supporting Parallel Fetch and Decode", in *Proceedings of the International Conference on Compilers, Architecture, and Synthesis for Embedded Systems* (CASES 2001), Atlanta, GA, November 2001, pp.168-175.

[11] S.G. Chandar, M. Mehendale, & R. Govindarajan. "Area and power reduction of embedded DSP systems using instruction compression and re-configurable encoding" *ICCAD 2001. IEEE/ACM International Conference on Computer Aided Design* , 4-8 Nov. 2001, pp.631 -634

[12] J.L. Turley. "Thumb Squeezes ARM Code Size", *Microprocessor Report*, 9, (4), 27 March 1995, pp. 1-5.

[13] A. Wolfe and A. Chanin, "Executing Compressed Programs on an Embedded RISC Architecture", in *Proceedings of the 25th Annual International Symposium on Micro architecture*, MICRO´25, December 1992, pp.81-91.